\documentclass[12pt,titlepage]{article}

\begin{document}

\title{UNIFIED TREATMENT OF MIXED VECTOR-SCALAR SCREENED COULOMB POTENTIALS
FOR FERMIONS}
\date{}
\author{Luis B. Castro\thanks{%
E-mail address: benito@feg.unesp.br (L.B. Castro)} and Antonio S. de Castro%
\thanks{%
E-mail address: castro@pesquisador.cnpq.br (A.S. de Castro)} \\
\\
UNESP - Campus de Guaratinguet\'{a}\\
Departamento de F\'{\i}sica e Qu\'{\i}mica\\
12516-410 Guaratinguet\'{a} SP - Brasil }
\date{}
\maketitle
\begin{abstract}
The problem of a fermion subject to a general mixing of vector and scalar
screened Coulomb potentials in a two-dimensional world is analyzed and
quantization conditions are found.
\end{abstract}

\section{Introduction}

Although the vector Coulomb potential does not hold relativistic bound-state
solutions, its screened version ($\sim e^{-|x|/\lambda }$) is a genuine
binding potential and its solutions have been found for fermions.\cite{ada}
The problem has also been analyzed for scalar\cite{DIRACscalarscreened} and
pseudoscalar\cite{asc} couplings. The Klein-Gordon equation with vector,\cite%
{KGvectorscreened} \ scalar\cite{KGscalarscreened} and arbitrarily \ mixed
vector-scalar\cite{KGmixed} couplings has not been exempted. As has been
emphasized in Refs. [2] and [4], the solution of relativistic equations with
this sort of potential may be relevant in the study of pionic atoms, doped
Mott insulators, doped semiconductors, interaction between ions, quantum
dots surrounded by a dielectric or a conducting medium, protein structures,
etc.

In the present paper it is shown that the problem of a fermion under the
influence of a mixed vector-scalar screened Coulomb potential, except for
possible isolated energies, can be mapped into a Sturm-Liouville problem for
the upper component of the Dirac spinor with an effective asymmetric
Morse-like potential, or an effective screened Coulomb potential in
particular circumstances. In all of those circumstances, the quantization
conditions are obtained. Beyond its potential physical applications, this
sort of mixing shows to be a powerful tool to obtain a deeper insight about
the nature of the Dirac equation and its solutions.

\section{The Dirac equation with mixed vector-scalar potentials in a 1+1
dimension}

In the presence of time-independent vector and scalar potentials the 1+1
dimensional time-independent Dirac equation for a fermion of rest mass $m$
reads

\begin{equation}
\mathcal{H}\Psi =E\Psi ,\quad \mathcal{H}=c\sigma_{1} p+\sigma_{3} \left(
mc^{2}+V_{s}\right) +V_{v},  \label{1a}
\end{equation}

\noindent where $E$ is the energy of the fermion, $c$ is the velocity of
light and $p$ is the momentum operator. $\sigma _{1}$ and $\sigma _{3}$ are 2%
$\times $2 Pauli matrices and the vector and scalar potentials are given by $%
V_{v}$ and $V_{s}$, respectively. Introducing the unitary operator $U(\delta
)=\exp \left[ -i\left( \delta -\pi /2\right) \sigma _{1}/2\right] $,
\noindent with $-\pi /2\leq \delta \leq \pi /2$, the transform of the
Hamiltonian (\ref{1a}), $H=U\mathcal{H}U^{-1}$, takes the form

\begin{equation}
H=\sigma _{1}cp+\sigma _{2}\cos \delta \,\left( mc^{2}+V_{s}\right) -\sigma
_{3}\sin \delta \,\left( mc^{2}+V_{s}\right) +V_{v}.  \label{3}
\end{equation}

\noindent In terms of the upper ($\phi $) and lower ($\chi $) components of
the transform of the spinor $\Psi $ under the action of the operator $U$, $%
\psi =U\Psi $, the Dirac equation, choosing $V_{v}=V_{s}\sin \delta $, i.e.,
$|V_{s}|\geq $ $|V_{v}|$, becomes \noindent
\begin{eqnarray}
\hbar c\phi ^{\prime }-\cos \delta \,\left( mc^{2}+V_{s}\right) \phi &=&i
\left[ E+\sin \delta \,mc^{2}\right] \chi  \nonumber \\
\hbar c\chi ^{\prime }+\cos \delta \,\left( mc^{2}+V_{s}\right) \chi &=&i
\left[ E-\sin \delta \,\left( mc^{2}+2V_{s}\right) \right] \phi .  \label{6b}
\end{eqnarray}

\noindent

\noindent where the prime denotes differentiation with respect to $x$. Note
that the charge conjugation is put into practice by the simultaneous changes
$E\rightarrow -E$ and $\delta \rightarrow -\delta $ while changing the sign
of any one of the components of the spinor $\psi $. Taking advantage of this
symmetry we can restrict our attention to nonnegative values of $\delta $.
Using the expression for $\chi $ obtained from the first line of (\ref{6b}),
and inserting it into the second line, one arrives at the following
second-order differential equation for $\phi $:
\begin{equation}
-\frac{\hbar ^{2}}{2}\phi ^{\prime \prime }+\left( \frac{\cos ^{2}\delta }{%
2c^{2}}\,V_{s}^{2}+\frac{mc^{2}+E\sin \delta }{c^{2}}\,V_{s}+\frac{\hbar
\cos \delta }{2c}\,V_{s}^{\prime }-\frac{E^{2}-m^{2}c^{4}}{2c^{2}}\right)
\phi =0.  \label{8}
\end{equation}

\noindent Therefore, the solution of the relativistic problem is mapped into
a Sturm-Liouville problem for the upper component of the Dirac spinor. In
this way one can solve the Dirac problem by recurring to the solution of a
Schr\"{o}dinger-like problem. The solutions for $E=-\sin \delta \,mc^{2}$,
excluded from the Sturm-Liouville problem, can be obtained directly from the
Dirac equation (\ref{6b}).

\section{The mixed vector-scalar screened Coulomb potential}

Now let us focus our attention on a scalar potentials in the form
\begin{equation}
V_{s}=-\frac{\hbar cg}{2\lambda }\exp \left( -\frac{|x|}{\lambda }\right) ,
\label{12}
\end{equation}%
\noindent where the coupling constant, $g$, is a dimensionless real
parameter and $\lambda $, related to the range of the interaction, is a
positive parameter. The solution for $E=-\sin \delta \,mc^{2}$ is not
continuous at $x=0$ and should be discarded. For $E\neq -\sin \delta
\,mc^{2} $ the Sturm-Liouville problem transmutes into
\begin{equation}
-\frac{\hbar ^{2}}{2m}\,\phi _{\varepsilon }^{\prime \prime }+V_{\mathtt{eff}%
}^{\left( \varepsilon \right) }\,\phi _{\varepsilon }=E_{\mathtt{eff}}%
\mathtt{\,}\phi _{\varepsilon },  \label{14a}
\end{equation}%
where $E_{\mathtt{eff}}=\left( E^{2}-m^{2}c^{4}\right) /\left(
2mc^{2}\right) $ and
\begin{equation}
V_{\mathtt{eff}}^{\left( \varepsilon \right) }=V_{1}^{\left( \varepsilon
\right) }\exp \left( -\frac{|x|}{\lambda }\right) +V_{2}\exp \left( -2\frac{%
|x|}{\lambda }\right)  \label{13}
\end{equation}%
\noindent with
\begin{equation}
V_{1}^{\left( \varepsilon \right) }=-\frac{mc^{2}\lambda _{c}\,g}{2\lambda }%
\left( 1+\frac{E}{mc^{2}}\sin \delta -\frac{\lambda _{c}}{2\lambda }%
\,\varepsilon \cos \delta \right) ,\;\;\;V_{2}=\frac{mc^{2}\lambda
_{c}^{2}\,g^{2}}{8\lambda ^{2}}\cos ^{2}\delta ,  \label{130}
\end{equation}

\noindent where $\varepsilon $ stands for the sign function ($\varepsilon
=x/|x|$ for $x\neq 0$).

\subsection{\noindent The effective screened Coulomb potential ($V_{v}=\pm
V_{s}$)}

For this class of effective potential, the discrete spectrum arises when $%
V_{1}^{\left( \varepsilon \right) }<0$ and $V_{2}=0$, corresponding to $%
V_{v}=\pm V_{s}$. Bound-state solutions are feasible only if $g>0$. Defining
the dimensionless quantities
\begin{equation}
y=y_{0}\exp \left( -\frac{|x|}{2\lambda }\right) ,\quad y_{0}=2\sqrt{\frac{%
\lambda g}{\lambda _{c}}\left( 1\pm \frac{E}{mc^{2}}\right) },\quad \mu =%
\frac{2\lambda }{\lambda _{c}}\sqrt{1-\left( \frac{E}{mc^{2}}\right) ^{2}}
\nonumber
\end{equation}

\noindent and using (\ref{14a})-(\ref{130}) one obtains the differential
Bessel equation $y^{2}\phi ^{\prime \prime }+y\phi ^{\prime }+\left(
y^{2}-\mu ^{2}\right) \phi =0$, \ \noindent where the prime denotes
differentiation with respect to $y$. The solution finite at $y=0$ ($%
|x|=\infty $) is given by the Bessel function of the first kind and order $%
\mu $:\cite{abr} $\phi (y)=N_{\mu }\,J_{\mu }(y)$, \noindent where $N_{\mu }$
is a normalization constant. In fact, the normalizability of $\phi $ demands
that the integral $\int_{0}^{y_{0}}y^{-1}|J_{\mu }(y)|^{2}dy$ must be
convergent. Since $J_{\mu }(y)$ behaves as $y^{\mu }$ at the lower limit,
one can see that $\mu \geq 1/2$ so that square-integrable Dirac
eigenfunctions are allowed only if $\lambda \geq \lambda _{c}/4$. The
boundary conditions at $x=0$ ($y=y_{0}$) imply that $dJ_{\mu
}(y)/dy|_{y=y_{0}}=0$ for even states, and $J_{\mu }(y_{0})=0$ for odd
states. \noindent Since the Dirac eigenenergies are dependent on $\mu $ and $%
y_{0}$, it follows that those last equations are quantization conditions.

\subsection{\noindent The effective Morse-like potential ($V_{v}\neq \pm
V_{s}$)}

Let us define $z=z_{0}\exp \left( -\frac{|x|}{\lambda }\right)$ , $z_{0}=g\cos \delta$, and
\begin{equation}
\rho _{\varepsilon }=\frac{\lambda }{\lambda _{c}\cos \delta }\left( 1+\frac{%
E}{mc^{2}}\,\sin \delta -\frac{\lambda _{c}}{2\lambda }\varepsilon \cos
\delta \right) ,\qquad \nu =\frac{\lambda }{\lambda _{c}}\sqrt{1-\left(
\frac{E}{mc}\right) ^{2}},  \label{24a} \\
\nonumber
\end{equation}

\noindent so that $z\phi _{\varepsilon }^{\prime \prime }+\phi _{\varepsilon
}^{\prime }+\left( -\frac{z}{4}-\frac{\nu ^{2}}{z}+\rho _{\varepsilon
}\right) \phi _{\varepsilon }=0$. \noindent Now the prime denotes
differentiation with respect to $z$. Following the steps of Refs. [4] and
[5], we make the transformation $\phi _{\varepsilon }=z^{-1/2}\Phi
_{\varepsilon }$ to obtain the Whittaker equation:\cite{abr}
\begin{equation}
\Phi _{\varepsilon }^{\prime \prime }+\left( -\frac{1}{4}+\frac{\rho
_{\varepsilon }}{z}+\frac{1/4-\nu ^{2}}{z^{2}}\right) \Phi _{\varepsilon }=0,
\label{16a}
\end{equation}%
whose solution vanishing at the in\-fin\-i\-ty is written as $\Phi
_{\varepsilon }$$=N_{\varepsilon }\,z^{\nu +1/2}e^{-z/2}M(a_{\varepsilon
},b,z)$, where $N_{\varepsilon }$ is a normalization constant and $M$ is a
regular solution of the confluent hypergeometric equation (Kummer\'{}s
equation):\cite{abr}

\begin{equation}
\xi M^{\prime \prime }+(b-\xi )M^{\prime }-a_{\varepsilon }M=0\textrm{, with }%
a_{\varepsilon }=\nu +\frac{1}{2}-\rho _{\varepsilon }\textrm{ and }b=2\nu +1.
\label{19}
\end{equation}

\noindent

\noindent Now we are ready to write the physically acceptable solutions on
both sides of the $x-$axis by recurring to the symmetry $\phi _{\varepsilon
}(-x)\sim \phi _{-\varepsilon }(x)$. They are

\begin{eqnarray}
\phi &=&z^{\nu }e^{-z/2}\left[ \theta (-x)C^{\left( -\right)
}M(a_{-},b,z)+\theta (+x)C^{\left( +\right) }M(a_{+},b,z)\right]  \nonumber
\\
\chi &=&z^{\nu }e^{-z/2}\left[ \theta (-x)D^{\left( -\right)
}M(a_{+},b,z)+\theta (+x)D^{\left( +\right) }M(a_{-},b,z)\right] ,
\label{10}
\end{eqnarray}

\noindent where $C^{\left( \pm \right) }$ and $D^{\left( \pm \right) }$ are
normalization constants and $\theta (x)$ is the Heaviside function. The
continuity of the wavefunctions at $x=0$ plus the substitution of   (\ref{10}) into the Dirac equation (\ref{6b}), and
making use of a pair of recurrence formulas involving the solution of Kummers%
\'{}%
s equation,\cite{abr}  lead to the quantization
condition

\begin{equation}
\frac{M(a_{+}+1,b,z_{0})}{M(a_{+},b,z_{0})}=\sqrt{\frac{a_{+}-2\nu }{a_{+}}}.
\label{eq23}
\end{equation}

\section{Concluding remarks}

The quantization conditions for a general mixing of vector and scalar
screened Coulomb potentials in a two-dimensional world have been put forward
in a unified way. Of course, we have to analyze the nature of the spectra as
a function of the potential parameters. This task, including the complete
set of eigenvectors, will be reported elsewhere.

\section*{Acknowledgements}

This work was supported by CAPES, CNPq and FAPESP.

\end{document}